%Paper: 9204207
%From: manohar@sphal.ucsd.edu
%Date: Thu, 02 Apr 92 17:24:16 PST

%
% INSTRUCTIONS FOR TeXING PAPER
%
% uses macro file harvmac.tex
%
% There are two encapsulated postscript figures at the end of the file
% which should be named dmeson.ps and bmeson.ps respectively. The
% figures are automatically included in the paper by the \epsffile commands
% understood by some versions of TeX/dvips. If your version of
% TeX/dvips cannot do this, delete the line \input epsf at the beginning
% of the paper and the lines between \listfigs and \bye at the end of
% the tex file, and print the postscript figures separately.
%
\input epsf
\input harvmac
%
%%%%%%%%%%%%%%%%%%%%%%%%%%%%%%%%%%%%%%%%%%%%%%%%%%%%%%%%%%%%%%%%%%%%%%
%
%  UCSD macros to overwrite some of the definitions in harvmac.tex
%  (include after harvmac.tex)
%  last modified 4/92
%
%%%%%%%%%%%%%%%%%%%%%%%%%%%%%%%%%%%%%%%%%%%%%%%%%%%%%%%%%%%%%%%%%%%%%%%
%
% add dec lno3r printer
%
%%% DEC LNO3R
%\def\unredoffs{} \def\redoffs{\voffset=-.31truein\hoffset=-.465truein}
%\def\speclscape{\special{landscape}}
%
% modify the output routine for the little format
%
\ifx\answ\bigans\else\output={
  \almostshipout{\leftline{\vbox{\pagebody\makefootline}}}\advancepageno
}
\fi
%
%
% address
%

%
% grant numbers
%

%
% preprint number
%
\def\UCSD#1#2{\noindent#1\hfill #2%
\bigskip\supereject\global\hsize=\hsbody%
\footline={\hss\tenrm\folio\hss}}% restores pagenumbers
%
% abstract
%
\def\abstract#1{\centerline{\bf Abstract}\nobreak\medskip\nobreak\par #1}
%
%
% titlefont
%
%
\edef\tfontsize{ scaled\magstep3}
 \tfontsize  \tfontsize
 \tfontsize \font\titlei=cmmi10 \tfontsize
\font\titleis=cmmi7 \tfontsize \font\titleiss=cmmi5 \tfontsize
\font\titlesy=cmsy10 \tfontsize \font\titlesys=cmsy7 \tfontsize
\font\titlesyss=cmsy5 \tfontsize  \tfontsize
\skewchar\titlei='177 \skewchar\titleis='177 \skewchar\titleiss='177
\skewchar\titlesy='60 \skewchar\titlesys='60 \skewchar\titlesyss='60
%
%\def\titlefont{\def\rm{\fam0\titlerm}% switch to title font
%\textfont0=\titlerm \scriptfont0=\titlerms \scriptscriptfont0=\titlermss
%\textfont1=\titlei \scriptfont1=\titleis \scriptscriptfont1=\titleiss
%\textfont2=\titlesy \scriptfont2=\titlesys \scriptscriptfont2=\titlesyss
%\textfont\itfam=\titleit \def\it{\fam\itfam\titleit}\rm}
%
%
% math symbols
%
%---------------------------------------------------------------------
%
\def\inv{^{\raise.15ex\hbox{${\scriptscriptstyle -}$}\kern-.05em 1}}
  %prime
\def\lbar{{\lower.35ex\hbox{$\mathchar'26$}\mkern-10mu\lambda}} %lambda bar

%
%
% various slashed symbols
%
%
\def\slash#1{\rlap{$#1$}/} % slashes a character
\def\dsl{\,\raise.15ex\hbox{/}\mkern-13.5mu D} %this one can be subscripted
\def\delsl{\raise.15ex\hbox{/}\kern-.57em\partial}
\def\Ksl{\hbox{/\kern-.6000em\rm K}}
\def\Asl{\hbox{/\kern-.6500em \rm A}}
\def\Dsl{\hbox{/\kern-.6000em\rm D}} %roman D
\def\Qsl{\hbox{/\kern-.6000em\rm Q}}
\def\gradsl{\hbox{/\kern-.6500em$\nabla$}}
%
% space and backspace in l mode
%
\def\lspace{\ifx\answ\bigans{}\else\qquad\fi}
\def\lbspace{\ifx\answ\bigans{}\else\hskip-.2in\fi} % $$\lbspace...$$
%
%     boxes an equation
%
\def\boxeqn#1{\vcenter{\vbox{\hrule\hbox{\vrule\kern3pt\vbox{\kern3pt
        \hbox{${\displaystyle #1}$}\kern3pt}\kern3pt\vrule}\hrule}}}
%
%     draw a little box (end of proof symbol)
%     e.g. \mbox{.1}{.1}
%
\def\mbox#1#2{\vcenter{\hrule \hbox{\vrule height#2in
\kern#1in \vrule} \hrule}}
%
%
%
%     curly letters
%
   %curly letters
  \def\CC{{\cal C}}

  \def\CO{{\cal O}}

%
%
%
%     derivatives
%
%

%

\def\bar#1{\overline{#1}}

\def\bra#1{\left\langle #1\right|}
\def\ket#1{\left| #1\right\rangle}
\def\abs#1{\left| #1\right|}

\def\darr#1{\raise1.5ex\hbox{$\leftrightarrow$}\mkern-16.5mu #1}

%
 %pound sterling
%
 %puts a small half in a displayed eqn
\def\frac#1#2{{\textstyle{#1\over #2}}} %puts a small fraction
%in a displayed eqn
%
%
%     various math operators
%
%

\def\Tr{\mathop{\rm Tr}}

%
%
%
%

%
%       relations
%
\def\ltap{\ \raise.3ex\hbox{$<$\kern-.75em\lower1ex\hbox{$\sim$}}\ }
\def\gtap{\ \raise.3ex\hbox{$>$\kern-.75em\lower1ex\hbox{$\sim$}}\ }
\def\gl{\ \raise.5ex\hbox{$>$}\kern-.8em\lower.5ex\hbox{$<$}\ }
\def\roughly#1{\raise.3ex\hbox{$#1$\kern-.75em\lower1ex\hbox{$\sim$}}}
%
%
%       This defines et al., i.e., e.g., cf., etc.

%

%
\def\np#1#2#3{{Nucl. Phys. } B{#1} (#2) #3}
\def\pl#1#2#3{{Phys. Lett. } {#1}B (#2) #3}

\def\physrev#1#2#3{{Phys. Rev. } {#1} (#2) #3}

\relax

\noblackbox

\centerline{{\titlefont Chiral Perturbation Theory
for}}
\medskip
\centerline{{\titlefont $f_{D_S}/ f_D$ and $B_{B_S}/B_{B}$}}
\bigskip
\centerline{Benjamin Grinstein}\medskip
\centerline{{\sl SSC Laboratory, 2250 Beckleymeade Avenue}}
\centerline{{\sl Dallas, TX 75237-3946}}
\bigskip
\centerline{Elizabeth Jenkins, Aneesh V. Manohar,
and Martin J. Savage}\medskip
\centerline{{\sl Department of Physics 0319, University of California,
San Diego}}
\centerline{{\sl La Jolla, CA 92093-0319}}
\bigskip
\centerline{Mark B. Wise} \medskip
\centerline{{\sl California Institute of Technology, Pasadena, CA 91125}}
\vfill
\abstract{The decay constants for the $D$ and $D_S$
mesons, denoted $f_D$ and $f_{D_S}$ respectively,
are equal in the $SU(3)_V$ limit, as
are the hadronic amplitudes for $B_S-\bar B_S$ and $B^0-\bar B^0$ mixing.
The leading $SU(3)_V$ violating
contribution to $\left( f_{D_S} / f_D \right)$ and
to the ratio of hadronic matrix elements relevant for
$B_S-\bar B_S$ and $B^0-\bar B^0$ mixing amplitudes
are calculated in chiral perturbation theory. We discuss the formalism
needed to include both meson and anti-meson fields in the heavy quark
effective theory.}
\vfill
%\draftmode
\UCSD{\vbox{\hbox{UCSD/PTH 92-05}\hbox{CALT-68-1768}
\hbox{SSCL-Preprint-25}}}{January 1992}

The decay constants for the $D$ and $D_S$ mesons
are defined by
\eqn\fd{
\bra{ 0 } \bar d \gamma_{\mu} \gamma_5 c \ket{ D^+(v)}
= i f_D p_{\mu},
}
and
\eqn\fs{
\bra{ 0 } \bar s \gamma_{\mu} \gamma_5 c \ket{ D_S(v)}
= i f_{D_S} p_{\mu}.
}
These decay constants are likely to be measured in the future using
the leptonic decays $D^+ \rightarrow  \mu^+ \nu_{\mu}$
and $D_S \rightarrow \mu^+ \nu_{\mu}$.  In the
chiral limit, where the up, down and strange quark masses
go to zero, flavor $SU(3)_V$ is an exact symmetry and so
$f_{D_S} / f_D = 1$.  However in nature, where $m_s \neq 0$,
this ratio will deviate from unity.  Neglecting the up
and down quark masses, in comparison with the strange
quark mass, this deviation has the form
\eqn\fratio{f_{D_S} / f_D =
1 + \kappa {M_K^2 \over {16\pi^2 f^2}}\ln\left(M_K^2/ \mu^2 \right)
+ \lambda(\mu) M_K^2 + ...}
where the ellipsis denote terms with more powers of the strange
quark mass (recall $M_K^2 \sim m_s$).  The dependence of $\lambda$
on the subtraction point $\mu$ cancels that of the
logarithm \ref\mudep{S. Weinberg, Physica 96A (1979) 327.}.
If $\mu$ is of order the chiral symmetry breaking scale then
$\lambda$ has no large logarithms and for very small $m_s$ the
term proportional to $\kappa$ dominates the deviation of
$f_{D_S} / f_D$ from unity.  Here we compute this logarithmic
correction.  Of course, in nature, the strange quark mass is
not small enough to justify the neglect of the term proportional
to $\lambda$.  However the logarithmic correction is interesting
for two reasons.  Firstly, as we have already mentioned, in chiral
perturbation theory it is formally the leading contribution to the
deviation of $f_{D_S} / f_D$ from unity.  Secondly, for the pion
and kaon decay constants \ref\fpik{P.~Langacker and H.~Pagels,
\physrev{D10}{1974}{2904}.
},
the analogous logarithmic term gives the
correct sign for $\left( f_K / f_{\pi} -1 \right)$. The magnitude,
however, is
too small by about a factor of two.

$B^0-\bar B^0$ mixing and $B_S-\bar B_S$ mixing give valuable
information on elements of the Cabibo-Kobayashi-Maskawa matrix.
One approach is to measure both these mixings, and then extract
$\abs{V_{td}/V_{ts}}^2$ from their ratio. This method has the
advantage that in the $SU(3)_V$ symmetry limit all dependence
on non-perturbative hadronic matrix elements cancels out.
(However, because $B_S-\bar B_S$ mass mixing is large, it will be very
difficult to measure.)
The hadronic matrix elements needed for the analysis of $B-\bar B$
mixing are
\eqn\bddef{
{8\over 3} f_{B}^2 B_{B} =
\bra{\bar B(v)} \bar b \gamma^\mu (1-\gamma_5) d
\ \bar b \gamma^\mu (1-\gamma_5) d \ket{B(v)},
}
\eqn\bsdef{
{8\over 3} f_{B_S}^2 B_{B_S} =
\bra{\bar B_S(v)} \bar b \gamma^\mu (1-\gamma_5) s
\ \bar b \gamma^\mu (1-\gamma_5) s \ket{ B_S(v)},
}
where the decay constants for the $B$ meson system are
defined by equations analogous to Eqs.~\fd\ and \fs\ for the
$D$ meson system. The parameters
$B_{B_S}$ and $B_{B}$ defined by Eqs.~\bddef\ and \bsdef\
are equal in the $SU(3)_V$ symmetry limit. For non-zero strange quark
mass, the ratio is no longer unity, and can be written in
the form Eq.~\fratio. The logarithmic term will be computed
in this paper using chiral perturbation theory. It is convenient
to perform the computation for the ratio of $B_B$'s, rather than for
the combination $B_B f_B^2$ that occurs in Eqs.~\bddef\ and \bsdef.
Most of the diagrams that occur for $B-\bar B$ mixing are the
same as those that occur in the computation of the decay
constants $f_B$, and can therefore be dropped in computing the
renormalization of $B_B$.

In Ref.~\ref\wise{M.B. Wise, CALT-68-1765 (1992)\semi
G. Burdman and J. Donoghue, UMHEP-365 (1992).}\
the formalism for applying chiral perturbation
theory to mesons containing a heavy quark was developed.  It
is important that the effective Lagrangian that describes the
low momentum interactions of the $D$ and $B$ mesons
with the pseudo-Goldstone bosons $\pi, K$ and $\eta$ be invariant
not only under chiral $SU(3)_L \times SU(3)_R$ symmetry but
also under heavy quark spin symmetry.  For example, even if one is
interested in processes involving only a real $D$ meson, the $D^*$
meson will occur as a virtual particle in Feynman diagrams.
The heavy quark symmetry causes the $D^*$ to be almost degenerate
with the $D$ so its effects cannot be neglected. It is also important to
write the chiral Lagrangian for matter fields such as the $D$'s in terms
of velocity dependent fields, to restore the validity of the chiral
expansion. The situation here is similar to the case of baryon chiral
perturbation theory \ref\baryon{H. Georgi, \pl{240}{1990}{447}\semi
E. Jenkins and A. Manohar,
\pl{255}{1991}{558}\semi See E. Jenkins and A. Manohar, Baryon Chiral
Perturbation Theory, UCSD/PTH 91-30
for a review.}.

The effective Lagrangian that describes the strong interactions
of the pseudo-Goldstone bosons with the ground state mesons
containing a heavy quark $Q$ is
\eqn\lag{\eqalign{
\cal L &= {f^2 \over 8}\Tr\ \left( \partial^{\mu} \Sigma
\partial_{\mu} \Sigma^\dagger \right)
+\lambda_0 \Tr\ \left[ m_q \Sigma + m_q \Sigma^\dagger \right]
-i \Tr \bar H^{(Q)a} v_{\mu} \partial^{\mu} H_a^{(Q)} \cr
&+{i \over 2} \Tr \bar H^{(Q)a} H_b^{(Q)} v^{\mu} \left[ \xi^\dagger
\partial_{\mu} \xi + \xi \partial_{\mu} \xi^\dagger \right]^b{}_a
+{{ig} \over 2} \Tr \bar H^{(Q)a} H_b^{(Q)} \gamma_{\nu} \gamma_5
\left[\xi^\dagger \partial^{\nu} \xi - \xi \partial^{\nu}
\xi^\dagger \right]^b{}_a \cr
&+ \lambda_1 \Tr \bar H^{(Q)a} H_b^{(Q)} \left[ \xi m_q \xi + \xi^\dagger
m_q \xi^\dagger \right]^b{}_a
+ \lambda_1^\prime \Tr \bar H^{(Q)a} H_a^{(Q)}\Tr\ \left[ m_q \Sigma
+ m_q \Sigma^\dagger \right] \cr
&+ {\lambda_2 \over m_Q} \Tr \bar H^{(Q)a} \sigma^{\mu \nu} H_a^{(Q)}
\sigma_{\mu \nu} +... \cr
}}
where the ellipsis denote terms with more derivatives, more
factors of the light quark mass matrix
\eqn\mass{ m_q = \pmatrix {m_u &0 &0 \cr
                           0 &m_d &0 \cr
                           0 &0 &m_s \cr}
}
or more factors of $1 / m_Q$ associated with the violation
of heavy quark spin symmetry.
The pseudoscalar and vector
meson fields $P_a^{(Q)}$ and $P^{*(Q)}_{a\mu}$ form the matrix \ref\georgi{H.
Georgi, Heavy Quark Effective Field Theory, TASI lectures,
HUTP-91-A039\semi J.D. Bjorken, SLAC-PUB-5278 (1990)\semi
A. Falk, H. Georgi, B. Grinstein and M.B. Wise, \np{343}{1990}{1}.}
$$H_a^{(Q)} = {{(1+\slash v)} \over 2} \left[ P^{*(Q)}_{a \mu} \gamma^{\mu}
- P_a^{(Q)} \gamma_5 \right]. $$
For $Q=c$, $(P_1^{(c)},P_2^{(c)},P_3^{(c)})=
(D^0,D^+,D_S^+)$, and similarly for $P^{*(c)}_{a\mu}$.
The field $H^{(Q)}_a$ is a doublet under the heavy quark spin symmetry, and a
$\bar 3$ under flavor $SU(3)_V$,
$$
H^{(Q)}_a \rightarrow S\ \left( H^{(Q)} \ U^\dagger\right)_a .
$$
The field $H^{(c)}$ describes $D$ and $D^*$ mesons with definite velocity
$v$. The subscript $v$ on $H$, $P$ and $P^*_\mu$ has been omitted,
to avoid complicating the notation.
The hermitian conjugate field is defined by
\eqn\hbardef{
\bar H^{a(Q)} = \gamma^0 H_a^{(Q)\dagger} \gamma^0 .
}
The pseudo-Goldstone bosons appear in the Lagrangian through
$\xi = e^{iM/f}$ ($\Sigma=\xi^2$) where
\eqn\pion{
M = \pmatrix{ {1\over\sqrt2}\pi^0 +
{1\over\sqrt6}\eta&
\pi^+ & K^+\cr
\pi^-& -{1\over\sqrt2}\pi^0 + {1\over\sqrt6}\eta&K^0\cr
K^- &\bar K^0 &- {2\over\sqrt6}\eta\cr
},
}
and the pion decay constant $f \simeq 135$~MeV. The Lagrangian Eq.~\lag\
is the most general Lagrangian invariant under both the heavy quark and
chiral symmetries to first order in $m_q$ and $1/m_Q$.

The left handed current
$L^{\nu}_a = \bar q_a \gamma^\nu (1 - \gamma_5) Q$ in QCD can be written
in the low energy chiral theory as \wise
\eqn\leftcurrent{
L^{\nu}_a = \left( {{i \alpha} \over 2} \right)
\Tr \gamma^\nu (1 - \gamma_5) H_b^{(Q)} \xi^{\dagger b}{}_a + ...,}
where the ellipsis denote higher dimension operators in the chiral and
heavy quark expansions. The parameter $f_P$ is obtained by taking the
matrix element of the current in the pseudoscalar meson state. At lowest
order, this fixes $\alpha=f_D\sqrt{m_D}$.
The graphs which
contribute at one loop are shown in \fig\graphs{The graphs contributing
to the renormalization of $f_P$. The solid square denotes the axial
vector current vertex.  The pseudoscalar and vector mesons $P$
and $P^*$ can be either the $D$ and $D^*$, or the $B$ and $B^*$.
Graph (a) is the
tree level contribution. Graph (b) is the wavefunction renormalization
correction and is proportional to $g^2$. Graph (c) vanishes identically.
Graph (d) is independent of~$g$.}.

The computation of the one loop graphs is straightforward.
Graph (c) vanishes because of the identity $v^{\mu}\left(g_{\mu\nu} -
v_\mu v_\nu\right)=0$.
The
renormalization of $f_{D_S}/f_D$ is independent of the overall magnitude
$\alpha$ of the current, and is
\eqn\answer{
f_{D_S} / f_D =
1 - {5\over 6}\left(1+3g^2\right)
{M_K^2 \over {16\pi^2 f^2}}\ln\left(M_K^2/ \mu^2 \right).
}
(The same formula also holds for $f_{B_S} / f_{B}$.)
The contribution of pion loops is proportional to $M_\pi^2 \ln
M_\pi^2$, and has been neglected.
The $\eta$ loops have been written in terms of $M_K$ using the
Gell-Mann--Okubo formula $M_\eta^2=4M_K^2/3$.
The one loop graphs with intermediate $P^*$ states depend on the
mass difference $\Delta=m_{P^*}-m_{P}$ only at order $\Delta^2$.
$\Delta$ has terms of order $1/m_Q$ as well as terms of order $m_s$,
and is numerically of order $M_\pi$. The $\Delta^2$ terms
are comparable to the $M_\pi^2$ terms, and can be neglected since they
are numerically small, and formally of higher order. This
simplifies the computation somewhat, since $\Delta$ can be set equal to
zero before evaluating the Feynman diagrams.

Numerically, the result is that
\eqn\numanswer{
f_{D_S} / f_D =
1 + 0.064\ (1+3g^2),
}
using $\mu=1$~GeV \ref\fourpi{A. Manohar and H. Georgi,
\np{234}{1984}{189}.}. The experimental limit on $\Gamma(D^*\rightarrow
D\pi)$ constrains $g^2$ to be less than~3.
The quark model estimate \ref\quarkestimate{N.~Isgur and M.B.~Wise,
\physrev{D41}{1990}{151}.}\ for $g$ is  $g^2\approx 0.7$
so that we expect $f_{D_S} / f_D \approx 1.2$. The formula~\answer\ can be
written in terms of either $f_\pi$ or $f_K$. Formally, this ambiguity
is of higher order, but it can make a sizeable difference in estimating
the correction, since $f_K=1.25 f_\pi$.
The most important terms in Eq.~\answer\ come from
virtual  $K$ mesons, so we have chosen to use $f_K$ to estimate the
correction.

The Lagrangian for $\bar B$ mesons is identical in form to Eq.~\lag,
except that the field $H^{(Q)}$ now has $Q=b$ and $P^{(b)}$ and
$P^{*(b)}_\mu$ destroy $\bar B$ and $\bar B^*$ mesons respectively.
In the heavy quark
effective theory, the field $P_v^{(Q)}$ destroys a meson of
velocity $v$ containing a heavy quark $Q$,
but it does not create an anti-meson containing the heavy anti-quark
$\bar Q$. To describe mesons containing heavy anti-quarks,
we have to introduce
two new fields, $P^{*(\bar Q)}_\mu$ and $P^{(\bar Q)}$ which destroy
mesons containing a heavy anti-quark $\bar Q$.
The phases of the fields $P^{*(\bar Q)}_\mu$ and $P^{(\bar Q)}$
relative to $P^{*(Q)}_\mu$ and $P^{(Q)}$ can be fixed using the
charge conjugation convention
\eqn\charge{\CC \xi \CC^{-1} = \xi^T, \qquad
 P^{*(\bar Q)a}_\mu  = -\CC P^{*(Q)}_{a\mu} \CC^{-1}, \qquad
 P^{(\bar Q)a}  =  \CC P_a^{(Q)} \CC^{-1}.
}
The field $H^{(\bar Q)a}$ is defined by
\eqn\htilde{
H^{(\bar Q)a} =  c  \left( \CC H_a^{(Q)} \CC^{-1} \right)^T c^{-1} =
 \left[  P^{*(\bar Q)a}_{\mu} \gamma^{\mu}
-  P^{(\bar Q)a} \gamma_5 \right]{{(1-\slash v)} \over 2}.
}
The matrix $c$ is the charge conjugation matrix for Dirac spinors,
$c= i \gamma^2\gamma^0$, and the transpose is on the spinor matrix indices.
The transpose and $c$ matrices are necessary to ensure that
$H^{(\bar Q)}$ transforms as a bispinor under the Lorentz group in the same
respresentation as $H^{(Q)}$. $H^{(\bar Q)}$
transforms as a $(\bar 2,3)$ under
the heavy spin $\otimes$ $SU(3)_V$ flavor symmetry,
$$
H^{(\bar Q)a} \rightarrow \left( U H^{(\bar Q)}\right)^a S^\dagger.
$$
The hermitian conjuate field is defined by
$$
\bar{H_a^{(\bar Q)}} = \gamma^0 H^{(\bar Q)a\dagger} \gamma^0.
$$
The Lagrangian for $B$ mesons in terms of $H^{(\bar b)}$ fields
is obtained from Eq.~\lag\ by setting $Q=b$ and applying charge conjugation.

The $\Delta b=2$ operator which produces $B-\bar B$ mixing in the
standard model is
$$
\CO^{aa} =\bar b \gamma_\mu (1-\gamma_5) q^a\
\bar b \gamma^\mu (1-\gamma_5) q^a ,
$$
where $a=2,3$ for $B^0$ and $B_S$ mixing respectively.
(Note that the repeated index $a$ is not summed.)
The operator $\CO^{aa}$ transforms as the 22 (or 33) component of the six
dimensional representation
of flavor $SU(3)_L$. In the effective theory, the operator $\CO^{aa}$
can be written as
\eqn\oeff{
\CO^{aa} =
\beta\ \Tr\ (\xi \bar H^{(b)})^a \gamma_\mu (1-\gamma_5) \ \Tr\
(\xi {H^{(\bar b)} })^a \gamma^\mu (1-\gamma_5)+\ldots.
}
Evaluating the traces gives
\eqn\oexpand{
\CO^{aa} = 4
 \beta \left[ (\xi P^{*(b)\dagger}_\mu)^a(\xi P^{*(\bar b)\mu})^a
+ (\xi P^{(b)\dagger})^a(\xi P^{(\bar b)})^a \right]+\ldots,
}
so that the amplitude for $B-\bar B$ mixing is the negative (since the
polarization of a physical $B^*$ is spacelike) of
that for $B^*-\bar B^*$ mixing.
This relation between the two
amplitudes can be proved directly by an application of the heavy
quark spin-symmetry. The operator $\CO^{aa}$ can potentially match onto
many different operators in the effective theory, such as
$\Tr\ (\xi \bar H^{(b)})^a \gamma_\mu (1-\gamma_5)
(\xi {H^{(\bar b)} })^a \gamma^\mu (1-\gamma_5)$.
However, all the
operators are proportional to Eq.~\oeff, because the spin symmetry
requires that the $B$ and $B^*$ mixing amplitudes be the negative of
each other. The $SU(3)_L\otimes SU(3)_R$ transformation property of
$\CO^{aa}$
then uniquely fixes the chiral structure of the operator.
The chiral corrections to $B_B$ are given by the graphs in
\fig\bfig{Graphs producing a renormalization of $B_B$. The dot is
the $\Delta b=2$ operator. Only virtual $\eta$ particles
contribute.}. Only $\eta$
graphs contribute to the correction to $B_{B_S}/B_B$. $K$ mesons
cannot contribute to the graphs in fig.~2a because of flavor
conservation. The $BB^*\pi$ coupling constant is the negative of the
$\bar B\, \bar B^*\pi$ coupling constant, because of the phase convention
for charge conjugation chosen in Eq.~\charge.
The two meson vertex in fig.~2b is obtained by
expanding Eq.~\oexpand\ to second order in the meson fields. The terms
where either $\xi$ is expanded to second order in the meson fields are
identical to those that occur in the renormalization of $f_B$, and can
be omitted. The term where each $\xi$ is expanded to first order in the
meson field is new. It has terms of the form
$\xi_{a}^a$,
and so only $\eta$ mesons contribute to fig.~2b. The chiral
correction to $B_{B_S}/B_B$ is
\eqn\bresult{
{B_{B_S} \over B_{B} } = 1 - {2\over 3}\left(1 - 3g^2\right)
{M_K^2 \over {16\pi^2 f^2}}\ln\left(M_K^2/ \mu^2 \right),
}
where we have again
used $M^2_{\eta}= 4 M^2_K/3$. Numerically, the correction is
$B_{B_S} / B_{B}\approx 0.9$, using $\mu=1$~GeV, $f=f_K$, and $g^2=0.7$
as before.
The renormalization of $B_{B_S} / B_{B}$ is a violation of
factorization in the hadronic matrix elements for $B-\bar B$ mixing.
(The perturbative QCD  corrections to
$B-\bar B$ mixing that contain large logarithms of $m_b/\Lambda_{\rm
QCD}$
do factorize  \ref\factorize{M.B. Voloshin and M.A.
Shifman, Sov. J. Nucl. Phys. 45 (1987) 292\semi H.D. Politzer and M.B.
Wise, \pl{206}{1988}{681}.}.)
The overall ratio of the hadronic
matrix elements for $B-\bar B$ mixing is obtained by combining
Eqs.~\answer\ and \bresult,
\eqn\combined{
{B_{B_S} f^2_{B_S}\over B_{B} f^2_{{B}} } = 1 - \left({7\over 3}
+ 3g^2\right)
{M_K^2 \over {16\pi^2 f^2}}\ln\left(M_K^2/ \mu^2 \right),
}
which is numerically about~1.3 for $\mu=1$~GeV, $f=f_K$, and $g^2\approx
0.7$.

Our results may be useful in estimating the difference between the values
of $f_{D_S}/f_D$ and $B_{B_S}/B_{B}$
in the quenched approximation to QCD and their values in
nature \ref\lattice{S. Sharpe, Nucl. Phys. B (Proc. Suppl.) 17 (1990)
146.}. The logarithmic corrections we have calculated necessarily involve
quark loops, and so would not be seen by lattice Monte Carlo
calculations that use the quenched approximation \lattice
\ref\bernard{C. Bernard, C.M.~Heard, J.~Labrenz and A.~Soni, Proceedings
of Lattice 91, Tsukuba, Japan (1991)\semi
A.~Abada, C.R.~Allton, Ph.~Boucaud, D.B.~Carpenter, M.~Cristafulli,
J.~Galand, S.~Gi\"usker, G.~Martinelli, O.~P\`ene, C.T.~Sachrajda,
R.~Sarno, K.~Schilling, and R.~Sommer, CERN-TH 6271/91 (1991).}.

\bigskip\bigskip
\centerline{\bf Acknowledgements}\nobreak
M.S. would like to thank Lisa Randall for discussions.
This work was supported in part by the Department of Energy under grant
\#DE-FG03-90ER40546, and contracts \#DEAC-03-81ER40050 and
\#DE-AC35-89ER40486.
B.G.~was supported in part by the Alfred P.~Sloan Foundation.
A.M.~was supported in part by the National Science Foundation by
Presidential Young Investigator award \#PHY-8958081.

\listrefs
\listfigs
\midinsert
\centerline{\hbox{\epsffile{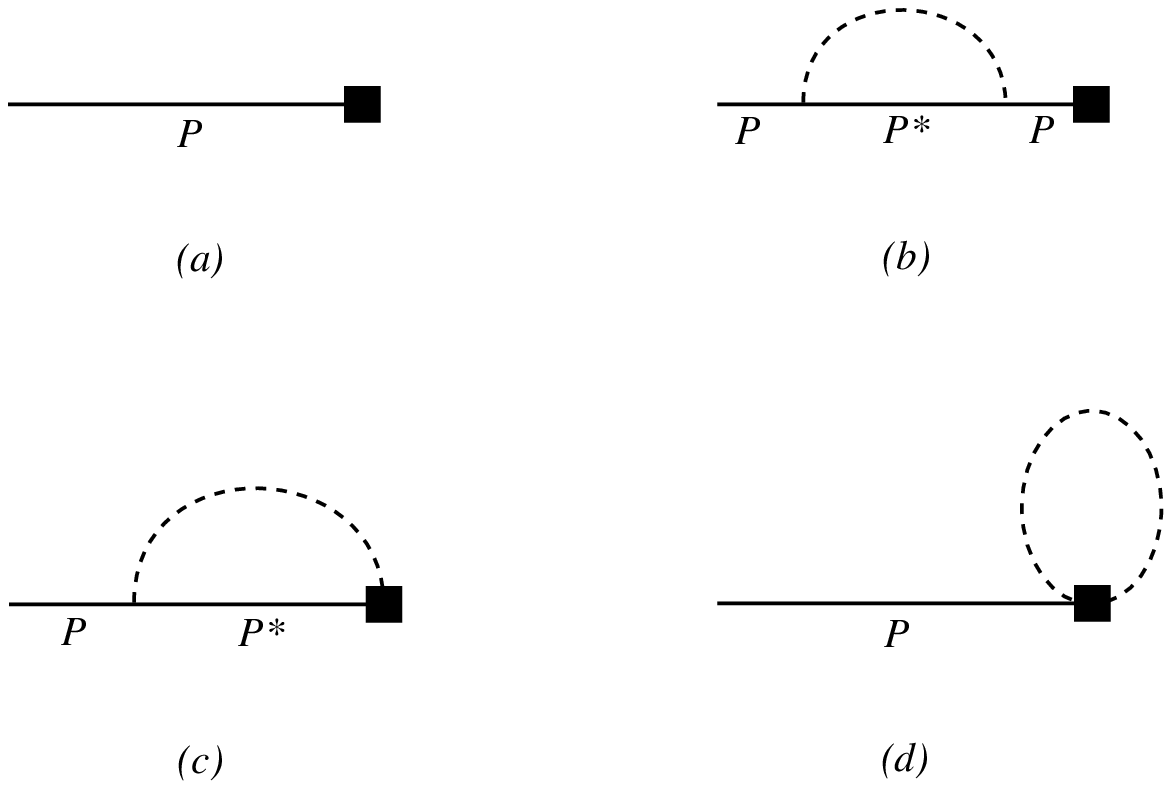}}}
\endinsert
\centerline{Figure 1}
\midinsert
\centerline{\hbox{\epsffile{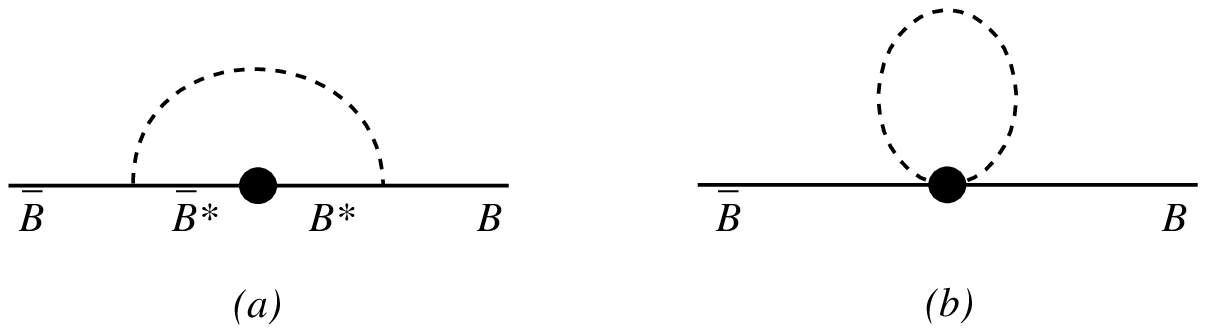}}}
\endinsert
\centerline{Figure 2}
\bye